%% file: main.tex
\documentclass[aps,prapplied,onecolumn,12pt]{revtex4-2}
\input{preamble}
\usepackage[a4paper,margin=2.54cm]{geometry}
\newcommand{\rev}[1]{\textcolor{red}{#1}}
\renewcommand{\rev}[1]{#1}

\begin{document}
\title{Short-Range Solvent--Solvent and Ion--Solvent Correlations at Metal--Electrolyte Interfaces: Parameterization and Benchmarking}

\author{Mengke Zhang}
\email{menkez@mail.ustc.edu.cn}
\affiliation{Institute of Energy Technologies, IET-3: Theory and Computation of Energy Materials, Forschungszentrum Jülich GmbH, 52425 Jülich, Germany}

\author{Jun Huang}
\email{ju.huang@fz.juelich.de}
\affiliation{Institute of Energy Technologies, IET-3: Theory and Computation of Energy Materials, Forschungszentrum Jülich GmbH, 52425 Jülich, Germany}
\affiliation{Theory of Electrocatalytic Interfaces, Faculty of Georesources and Materials Engineering, RWTH Aachen University, 52062 Aachen, Germany}

\date{\today}

\begin{abstract}
	Short-range correlations in electrolyte solutions lead to oscillatory profiles of water polarization and ionic concentration at electrode–electrolyte interfaces. The recently developed density–potential–polarization functional theory (DPPFT) provides a comprehensive framework to incorporate these short-range correlation effects. In the present work, the parameters describing short-range solvent--solvent and ion--solvent correlations in DPPFT are determined from the wavenumber-dependent dielectric susceptibility spectrum of pure water and from ion solvation energies derived within nonlocal electrostatics, respectively. The experimental ionic-radius-dependent hydration energies of alkali metal cations and halide anions are well reproduced by the solvation model. The charge hydration asymmetry is explained as the stronger short-range repulsion between cations and water molecules compared to that between anions and water molecules. Using these parameters, DPPFT is then applied to describe short-range correlation effects at the Ag(111)--NaF aqueous electrolyte interface. The water polarization profiles obtained from DPPFT calculations agree with AIMD simulations. Furthermore, as the strength of short-range ion–solvent repulsion increases, the peaks of anionic/cationic layers shift from regions near centers of positive/negative polarization charges toward those of opposite sign, thereby preserving solvation configurations similar to those in bulk solution. \rev{This work develops a procedure for parameterizing short-range correlation effects within the DPPFT framework}, thereby enabling a more quantitative and computationally efficient description of atomic-scale phenomena at electrochemical interfaces.   
\end{abstract}

\maketitle
\section{Introduction}
Understanding the interfacial structure between electrode materials and aqueous electrolyte solutions is of great relevance to the advancement of electrochemical devices such as fuel cells, batteries, and electrolyzers. Short-range correlations within electrolyte solutions are closely linked to various interfacial phenomena at the atomic scale, such as oscillatory profiles of electric potential, co-ions, and counter-ions, which have been observed across a wide range of concentrations, from dilute\cite{martin2016atomically} to moderately concentrated\cite{di2023constant} and highly concentrated solutions\cite{philpott1995screening,spohr1999molecular}. While molecular dynamics (MD) simulations can accurately capture atomistic details, a continuum theory incorporating short-range correlations effects is needed for large-scale modeling.

The effects of short-range correlations between solvent molecules can be taken into account through nonlocal electrostatics\cite{kornyshev1981nonlocaldoublelayer,kornyshev1982nonlocal,kornyshev1981nonlocal,ulstrup1979multiphonon,kornyshev1983non}, wherein the dielectric susceptibility becomes wavevector-dependent in Fourier space, giving rise to its spatial dispersion spectrum (Unless otherwise specified, polarization and dielectric response here and below refer to their longitudinal components). A commonly used form of the susceptibility spectrum is the Lorentzian, which in real space corresponds to a Yukawa-type susceptibility that decays over the correlation length. The Lorentzian form of the susceptibility spectrum is equivalent to the second-order Landau--Ginzburg (LG) functional of the corresponding polarization mode, as used in the Mar\v{c}elja–Radi\v{c} model\cite{marvcelja1976repulsion}. \rev{An oscillatory susceptibility has also been considered to describe ion solvation\cite{basilevsky1998nonlocal}.} The inclusion of the nonlocal electrostatic effect within the continuum theory leads to a substantially improved consistency between the predicted solvation energies of alkali metal and halide ions and experimental values\cite{dogonadze1974polar,hildebrandt2004novel,ulstrup1979multiphonon}.

Recently, an extended fourth-order LG functional, with the orientational polarization as the functional variable, has been adopted to describe the short--range correlation energy between water molecules, with particular focus on the structure of water in nanoconfinement\cite{monet2021nonlocal} and in the vicinity of ions\cite{berthoumieux2018gaussian}. The wavenumber-dependent dielectric susceptibility derived from this extended LG functional captures the primary peak in the spatial dispersion spectrum of pure water at $k\approx 3\ \text{\AA}^{-1}$, as observed in inelastic neutron scattering experiments and MD simulations\cite{bopp1996static,bopp1998frequency,monet2021nonlocal,labavic2025nonlocal,becker2025dielectric}. Based on the LG functional description for short-range correlations between water molecules in the electrical double layer (EDL), polarization is introduced by Hedley et al.\cite{hedley2023dramatic} as a separate order parameter, in addition to the electric potential. This two-order-parameter EDL model delineates spatial oscillations in electric potential and ionic concentration, which arise from the polarized water layering induced by short-range correlations between water molecules\cite{hedley2023dramatic}.

Meanwhile, a field-theoretic model has been developed by Blossey et al. for structural electrolyte solutions\cite{blossey2022field,blossey2023comprehensive}. Beyond the LG-based description for solvent correlations in the EDL model by Hedley et al.\cite{hedley2023dramatic}, further physical aspects such as ion--solvent short-range correlations, nonlinear dielectric response, and constituent polarizability have also been accounted for in this field-theoretic approach. This field-theoretic model was later combined by us with an orbital-free density functional theory (DFT) of electrons on the metal side within a grand canonical framework\cite{zhang2025structured}, referred to as the density--potential--polarization functional theoretical (DPPFT) approach. DPPFT incorporates the electron spillover effects and eliminates the need to specify boundary conditions for the polarization at the metal surface. The double-layer capacitance ($C_\text{dl}$) simulated using this semi-classical EDL model shows improved agreement with experimental data at the Ag–aqueous KPF\textsubscript{6} solution interface compared to previous work\cite{huang2023density}, particularly in the enhanced capacitance magnitude at the potential of zero charge (PZC). \rev{This improvement is attributed to short-range solvent--solvent correlations that give rise to a layered solvent structure, leading to a reduced effective dielectric constant near the surface, as verified by experiments\cite{fumagalli2018anomalously} and MD simulations\cite{schlaich2016water,dinpajooh2016dielectric,mondal2021anomalous,ballenegger2005dielectric}.}

In the field-theoretic approach to structured electrolyte solutions \cite{blossey2022field,blossey2023comprehensive,zhang2025structured}, the solvent orientational polarization $\mathbf{P}$ is influenced by both the electrostatic field and an auxiliary field $\bm{\mathcal{E}}$, as described by a modified Langevin polarization equation, i.e.,
\begin{equation}
    \mathbf{P} = - \frac{p n_s \mathscr{L}(\beta p |\bm{\mathcal{E}} + \nabla \phi|)}{|\bm{\mathcal{E}} + \nabla \phi|} (\bm{\mathcal{E}} + \nabla \phi), \label{equ: modified Langevin polarization equation}
\end{equation}
where $\beta=1/(k_B T)$ is the thermal factor, $\phi$ is the electrostatic potential, $n_s$ is the number density of solvent molecules, and $\mathscr{L}(x)=\coth(x)-1/x$ denotes the Langevin function. The solvent dipole moment has a magnitude of $p$, and here, its direction is conventionally defined from the negative to the positive charge center, in contrast to the definition adopted in the previous work\cite{blossey2022field,blossey2023comprehensive,zhang2025structured,blossey2022continuum}. The auxiliary field $\bm{\mathcal{E}}$ is expressed as\cite{zhang2025structured}
\begin{align}
    \bm{\mathcal{E}} = \frac{1}{\epsilon_0} \left( K_s \mathbf{P} - K_\alpha \nabla^2 \mathbf{P} + K_\beta \nabla^4 \mathbf{P} \right) 
    + \alpha_c e_0 \nabla n_c - \alpha_a e_0 \nabla n_a, \label{equ: controlling equation of polarization}
\end{align}
where $\epsilon_0$ is the vacuum permittivity, $e_0$ is the elementary charge, and $n_a$ and $n_c$ are number densities of anions and cations, respectively. $K_s$, $K_\alpha$, and $K_\beta$ are expansion coefficients in the LG functional. \rev{The $K_s$ term represents a local short-range correlation mode, while the $K_\alpha$ and $K_\beta$ terms correspond to two nonlocal modes characterized by the characteristic lengths $\abs{K_\alpha}^{1/2}$ and $\abs{K_\beta}^{1/4}$, respectively.} $\alpha_c$ and $\alpha_a$ are the leading-order expansion coefficients of the short-range correlation energy between ions and solvent molecules, characterizing the strength of short-range cation--solvent and anion--solvent correlations, respectively. \rev{The corresponding correlation lengths measured from the cationic and anionic charges are given by $\abs{\alpha_c\epsilon_\text{ir}}^{1/2}$ and $\abs{\alpha_a\epsilon_\text{ir}}^{1/2}$, respectively.} Here, the infrared dielectric permittivity $\epsilon_\text{ir}$ accounts for the non-orientational polarization modes of the medium, including electronic and atomic polarizations. 

The description of solvent dielectric polarization behaviors is a central aspect of continuum dielectric theory. As seen in Eq.~\eqref{equ: modified Langevin polarization equation}, the auxiliary field $\bm{\mathcal{E}}$
is a non-electrostatic field that represents the averaged influence of short-range correlations on the solvent orientational polarization. \rev{Clearly}, when all short-range correlation parameters vanishes, $\bm{\mathcal{E}}=0$, and Eq.~\eqref{equ: modified Langevin polarization equation} reduces to the classical Langevin polarization model\cite{gongadze2014ions,gongadze2015asymmetric}, which describes the orientations of non-interacting permanent dipoles to a local electrostatic field. Therefore, determining the short-range correlation parameters in Eq.~\eqref{equ: controlling equation of polarization} is a necessary first step toward quantifying this short-range correlation effect. Once this is done, combining Eq.~\eqref{equ: modified Langevin polarization equation} and Eq.~\eqref{equ: controlling equation of polarization} with the Poisson equation enables us to quantitatively describe short-range correlation effects in the electrolyte solution, or, in the DPPFT approach, to consider the short-range correlation effect on the interfacial structure and properties of the whole interface. To this end, in the following, we consider the nonlocal electrostatics of pure water and ion solvation, respectively, to determine the short-range correlation parameters between water molecules and between ions and water, and finally examine the short-range correlation effects on the interfacial structure between the metal and the aqueous electrolyte solution in a DPPFT framework.

\section{Pure water}\label{sec: pure water}
The effect of short-range correlations between water molecules, which are associated with hydrogen-bonding networks\cite{hedley2023dramatic,berthoumieux2019dielectric}, is reflected as previously mentioned in the wavenumber dependence of the dielectric susceptibility. The electric susceptibility is related to the structure factor, which is the correlation function of the solvent longitudinal polarization, through the fluctuation–dissipation theorem\cite{bopp1998frequency}. For water, the structure factor can be decomposed into partial HH, OH, and OO structure factors, which can be obtained through neutron scattering techniques\cite{bopp1996static} or molecular dynamics (MD) simulations\cite{bopp1998frequency}, assuming a localized charge distribution on the atoms. In the linear response regime relevant to the spectrum, Eq.~\eqref{equ: modified Langevin polarization equation} reduces to its linearized form:
\begin{equation}
    \mathbf{P} \approx -\frac{1}{3}\beta p^2 n_s^b(\bm{\mathcal{E}}+\nabla\phi). \label{equ: polarization equation in linear response range}
\end{equation}
Combining the above equation and Eq.~\eqref{equ: controlling equation of polarization} with $\alpha_c$ and $\alpha_a$ being zero (pure water system), the relationship between the electric displacement $\mathbf{D}$ and the orientational polarization of water $\mathbf{P}$ is given by
\begin{align}
    \mathbf{D} = \epsilon_\text{ir} \mathbf{E} + \mathbf{P} = \left(1+\varepsilon_\text{ir} K_s + \frac{3\varepsilon_\text{ir}\epsilon_0}{\beta p^2 n_s} \right) \mathbf{P} - \varepsilon_\text{ir} K_\alpha \nabla^2 \mathbf{P} + \varepsilon_\text{ir} K_\beta \nabla^4 \mathbf{P},
\end{align}
where $\varepsilon_\text{ir}=\epsilon_\text{ir}/\epsilon_0$ denotes the infrared dielectric constant. Performing Fourier transform on the both sides of above equation, the spectrum of the longitudinal dielectric susceptibility of orientational polarization is obtained as\cite{monet2021nonlocal}
\begin{equation}
    \chi(k) = \frac{1}{\chi^{-1}_0 + \varepsilon_\text{ir} K_\alpha k^2 + \varepsilon_\text{ir} K_\beta k^4}, \label{equ: longitudinal susceptibility spectrum}
\end{equation}
with the reciprocal of $\chi(k)$ at $k=0$,
\begin{equation}
    \chi^{-1}_0 = 1+\varepsilon_\text{ir} K_s + \frac{3\varepsilon_\text{ir}\epsilon_0}{\beta p^2 n_s}, \label{equ: chi0-1}
\end{equation}
where $\chi_0$ is the static dielectric susceptibility of the orientational polarization, which is related to the static dielectric constant of water $\varepsilon_s$ and $\varepsilon_\text{ir}$ by $\chi_0 = 1 - \varepsilon_\text{ir}/\varepsilon_s$. For water, based on its vibrational dispersion spectrum\cite{kuznetsov1999electron,kornyshev1996shape}, the values of 4.9 and 78.5 are used for $\varepsilon_\text{ir}$ and $\varepsilon_s$, respectively, which gives a value of 0.9376 for $\chi_0$.

From Eq.~\eqref{equ: chi0-1}, the parameter $K_s$ can be determined from bulk properties of the solvent, such as the dipole moment, number density, static dielectric constant, as follows:
\begin{equation}
    K_s = \frac{1}{\varepsilon_s-\varepsilon_\text{ir}} - \frac{3\epsilon_0}{\beta p^2 n_s}. \label{equ: Ks}
\end{equation}
The magnitude of $K_s$ reflects the extent to which short-range correlations contribute to the local dielectric response. The larger effective dipole moment estimated from the classical Langevin polarization model can be attributed to this contribution. Another factor contributing to the enhanced effective dipole moment is the cavity field effect. Due to dielectric screening, a water molecule experiences a local electric field stronger than the macroscopic field, which can notably modify the effective dipole moment at highly charged surfaces\cite{gongadze2012decrease}.

For a negative $K_\alpha$ and a positive $K_\beta$, \rev{corresponding to attractive and repulsive short-range correlation modes between solvent molecules}, Eq.~\eqref{equ: chi} attains its maximum at the wavenumber $k_\text{max}=\left(-K_\alpha/(2K_\beta) \right)^{\frac{1}{2}}$ with the corresponding maximum value $\chi_\text{max}$. Inversely, $K_\alpha$ and $K_\beta$ can be determined from $k_\text{max}$ and $\chi_\text{max}$:
\begin{align}
    K_\beta = \frac{\chi_0^{-1}-\chi_\text{max}^{-1} }{\varepsilon_\text{IR}k_\text{max}^4},\quad K_\alpha = -2K_\beta k_\text{max}^2 \label{equ: K_alpha, K_beta}.
\end{align}
 The dielectric susceptibility spectrum, whether measured experimentally or simulated using MD, may arise not only from orientational polarization but also from atomic polarization $\chi_\text{atomic}$ and electronic polarization $\chi_\text{el}$ of the solvent. We assume that only the dielectric susceptibility associated with solvent orientations is wavenumber-dependent. The total dielectric susceptibility $\chi_\text{tot}$ is then the sum of contributions from different polarization modes,
\begin{equation}
    \chi_\text{tot}(k) = \chi_\text{el} + \chi_\text{atomic} + \chi(k) = \frac{\varepsilon_\text{ir}-1}{\varepsilon_s} + \chi(k). \label{equ: relation between chi and chi_tot}
\end{equation}
The experimentally measured spectrum of electric susceptibility exhibits a broad overscreening zone, peaking at $k_\text{max}=3\ \text{\AA}^{-1}$ with a maximum value of approximately 40\cite{monet2021nonlocal}. As shown in Eq.~\eqref{equ: relation between chi and chi_tot}, $\chi(k)$ differs from $\chi_\text{tot}$ by only 0.05 for $\varepsilon_\text{ir}=4.9$ and $\varepsilon_s=78.5$. This discrepancy is negligible compared with the dielectric susceptibility in the overscreening region, where $\chi_\text{tot}>1$, and can therefore be safely ignored. In this case, $K_\alpha$ and $K_\beta$ in Eq.~\eqref{equ: K_alpha, K_beta} are determined using the values of $k_\text{max}$ and $\chi_\text{max}$ as $-0.047$ $\text{\AA}^2$ and $0.0026$ $\text{\AA}^{4}$, respectively.

These short-range correlation parameters can be related to the structural properties of water through its static structure factor, which corresponds to the real-space correlation function of the longitudinal component of the polarization. Assuming purely classical orientations of solvent molecules at constant temperature, this correlation function is proportional to the orientational dielectric susceptibility by a constant factor\cite{monet2021nonlocal}. The orientational dielectric susceptibility $\chi(r)$ is obtained by performing an inverse Fourier transform on Eq.~\eqref{equ: longitudinal susceptibility spectrum} as\cite{supplemental}:
\begin{align}
    \chi(r) = \frac{\sin(ar) e^{-br}}{2\pi \varepsilon_\text{ir} Xr}, \label{equ: chi}
\end{align}
provided that the following inequality is satisfied:
\begin{equation}
    X^2 = K_\beta \gamma_0 - K_\alpha^2 > 0,\ \text{with}\ \gamma_0 = 4(\chi_0\varepsilon_\text{ir})^{-1}.
\end{equation}
Accordingly, the correlation function of the orientational polarization therefore exhibits a periodic length $\lambda_\text{p}$ and a decay length $\lambda_\text{d}$, which can be derived as\cite{monet2021nonlocal,supplemental}:
\begin{align}
    \lambda_\text{p} &= \frac{2\pi}{a} = 2\pi \sqrt{\frac{4 K_\beta}{-K_\alpha+\sqrt{K_\beta \gamma_0}}} = 2\pi k_\text{max}^{-1} \sqrt{\frac{2}{1+\sqrt{\frac{1}{1-\chi_0\chi_\text{max}^{-1}}}} }, \label{equ: periodic length} \\
    \lambda_\text{d} &= \frac{1}{b} = \sqrt{\frac{4 K_\beta}{K_\alpha+\sqrt{K_\beta \gamma_0}}} = k_\text{max}^{-1} \sqrt{\frac{2}{-1+\sqrt{\frac{1}{1-\chi_0\chi_\text{max}^{-1}}}} }. \label{equ: decay length}
\end{align}

Given the values of $k_\text{max}\ = \text{3\ \AA}^{-1}$, $\chi_0 = 0.9376$, and $\chi_\text{max} = 40$, the characteristic lengths $\lambda_\text{p}$ and $\lambda_\text{d}$ are calculated using the above equations as $\lambda_\text{p} = 2.1\ \text{\AA}$ and $\lambda_\text{d} = 4.3\ \text{\AA}$. The dielectric susceptibility underlying the LG approach in Eq.~\eqref{equ: chi} takes a Lorentzian form modified by an oscillatory term, which accounts for the emergence of oscillations in the polarization and electric potential in relevant models\cite{hedley2023dramatic,zhang2025structured}. At small wavenumbers, corresponding to the wavelength $\lambda \gg \lambda_\text{p}$, the structural oscillations of water molecules can not be resolved. Accordingly, the limit $a\rightarrow 0$ is taken, giving that $K_\alpha = \gamma_0/(2b^2)$ and $K_\beta = \gamma_0/(4b^4)$ (see Eq.~S9 in Ref.\cite{supplemental}). The longitudinal dielectric susceptibility in Eq.~\eqref{equ: chi} is reduced to the Lorentzian form:
\begin{equation}
    \chi(k) = \frac{1}{\chi^{-1}_0 + \frac{\varepsilon_\text{IR}\gamma_0}{2b^2}k^2 (1+\frac{k^2}{2b^2})} \xrightarrow{\text{small}\ k} \chi_\text{Lorentz}(k) = \frac{\chi_0}{1 + 2\lambda_\text{d}^2k^2}.
\end{equation}

\section{Ion solvation}\label{sec: ion solvation}
The short-range correlation parameters of water, $K_s$, $K_\alpha$, and $K_\beta$, can be determined using Eq.~\eqref{equ: Ks} and Eq.~\eqref{equ: K_alpha, K_beta}. When a single ion is immersed in water, its electric field polarizes nearby water molecules, leading to ion solvation. The short-range correlations between the ion and the solvent exert a non-electrostatic field effect on the surrounding medium, as described in Eq.~\eqref{equ: controlling equation of polarization}, and consequently influence the dielectric response of water. In the field-theoretic approach\cite{blossey2023comprehensive,zhang2025structured}, these ion-solvent short-range correlations are effectively described by a constant kernel, $\alpha_c$ or $\alpha_a$, which couples the ionic charge density and solvent polarization charge. Let $\rho(\mathbf{r})$ denote the ion charge density distribution; then Eq.~\eqref{equ: controlling equation of polarization} takes the form:
\begin{align}
    \bm{\mathcal{E}} &= \frac{1}{\epsilon_0} \left( K_s \mathbf{P} - K_\alpha \nabla^2 \mathbf{P} + K_\beta \nabla^4 \mathbf{P} \right) 
    + \alpha \nabla \rho(\mathbf{r}), \label{equ: controlling equation for solvation}
\end{align}
where the subscript on $\alpha$ is omitted. In principle, the ionic charge density distribution can be obtained from quantum mechanical calculations\cite{clementi1974roothaan}. Combining Eq.~\eqref{equ: controlling equation for solvation} and Eq.~\eqref{equ: modified Langevin polarization equation} with the Poisson equation, one can solve for the electric potential distribution of an ion in the solvent medium, $\phi(\mathbf{r})$, and in vacuum, $\phi_\text{vac}(\mathbf{r})$. The ion solvation energy, $\Delta G_\text{solv}$, is then given by the difference between the electrostatic energies in the solvent medium and in vacuum, i.e.,
\begin{equation}
    \Delta G_\text{solv} = \frac{1}{2}\int\text{d}\mathbf{r}\ \rho(\mathbf{r})[\phi(\mathbf{r})-\phi_\text{vac}(\mathbf{r})].
\end{equation}
The short-range correlation parameters of water, $K_s$, $K_\alpha$, and $K_\beta$, can be determined using Eq.~\eqref{equ: Ks} and Eq.~\eqref{equ: K_alpha, K_beta}. The only unknown parameter in $\Delta G_\text{sol}$ is $\alpha$, whose value can therefore be determined from the experimental value of solvation energy.

As a simplified case, we examine the solvation of a spherical ion in the linear response regime of the solvent polarization, such that Eq.~\eqref{equ: polarization equation in linear response range} holds. In this case, by combining Eqs.~\eqref{equ: polarization equation in linear response range} and \eqref{equ: controlling equation for solvation}, and the fundamental electrostatic relationship, $\nabla\cdot \mathbf{D}=\rho$, we have:
\begin{align}
    \mathbf{D} = \chi_0^{-1}\mathbf{P} - \varepsilon_\text{ir} K_\alpha \nabla^2\mathbf{P} + \varepsilon_\text{ir} K_\beta \nabla^4\mathbf{P} + \alpha\varepsilon_\text{ir}\epsilon_0 \nabla^2\mathbf{D}. \label{equ: ion solvation - electric displacement}
\end{align}
Performing a Fourier transform on both sides of the above equation, we obtain corresponding dielectric susceptibility spectrum, which is modified by short-range correlations between the ion and orientational polarization of the solvent,
\begin{equation}
    \chi_\alpha(k) = \frac{1 + \alpha\varepsilon_\text{ir}\epsilon_0 k^2}{\chi_0^{-1} + \varepsilon_\text{ir}K_\alpha k^2 + \varepsilon_\text{ir}K_\beta k^4}. \label{equ: chi_alpha}
\end{equation}
The ion solvation energy in the nonlocal dielectric medium is then given by Dogonadze-Kornyshev solvation equation\cite{dogonadze1974polar,supplemental},
\begin{equation}
    \Delta G_\text{solv} = -\frac{1}{4\pi^2\epsilon_0} \int_0^\infty\text{d}k\ \rho^2(k)\left(\frac{\varepsilon_\text{ir}-1}{\varepsilon_s}+\chi_\alpha(k)\right), \label{equ: solvation energy}
\end{equation}
where $\rho(k)$ is the Fourier transform of ionic charge density. \rev{The above equation clearly separates the solvation energy into contributions from fast polarization and slow orientational polarization modes. The solvation energy often reaches several electron volts and therefore has a significant impact on electron transfer reactions. First, the difference in the total solvation energy between the reactant and product states contributes to the thermodynamic driving force of the reaction. Second, only the slow orientational part of the solvation contributes to the reorganization energy, which modulates the activation barrier for electron transfer\cite{zhang2025electrochemical}.}        

To calculate the solvation energy, one must first prescribe the ionic charge density. The simplest representation is given by the Born sphere model, where the ionic charge $ze_0$ is taken to be uniformly distributed over the surface of a sphere of radius $R$, with $z$ denoting the ion valency. In this case, the ionic charge density in Fourier space is given by
\begin{equation}
	\rho(k)=ze_0 \frac{\sin(kR)}{kR}.
\end{equation}
When short-range correlation effects are neglected, i.e., $K_\alpha,K_\beta,\alpha=0$, so that $\chi_\alpha(k)=\chi_0$, Eq.~\eqref{equ: solvation energy} simplified to the Born solvation formula. However, the Born sphere model is incompatible with quantum mechanical calculations, where the ionic charge distribution is radially smeared rather than confined to a spherical surface. To account for the smeared charge effect, a convenient model for the ionic charge density is the stretched Gaussian of order $n$ (SGn), which in real space is given by\cite{kornyshev1997nonlocal}
\begin{equation}
    \rho_\text{SGn}(r) = ze_0\frac{2^n n^{\frac{2n+3}{2}}}{\pi^{\frac{3}{2}} (2n+1)!!R^{2n+3}} r^{2n} e^{-\frac{n r^2}{R^2}}, \label{equ: SGn model}
\end{equation}
while its Fourier transform reads
\begin{equation}
	\rho_\text{SGn}(k) = ze_0 \frac{(-1)^n 2^n n^{\frac{2n+3}{2}}}{(2n+1)!!R^{2n+3}} \frac{\partial^n}{\partial y^n}\Bigg(\frac{e^{-\frac{k^2}{4y}}}{y^{\frac{3}{2}}}\Bigg),
\end{equation}
where $y=n/(R^2)$. A smaller $n$ in the SGn model corresponds to a broader charge distribution around the spherical surface of radius $R$. For very large $n$, e.g., $n=50$, the distribution becomes sharply localized and approaches the Born sphere model.

\begin{figure}
    \centering
    \includegraphics[width=0.5\linewidth]{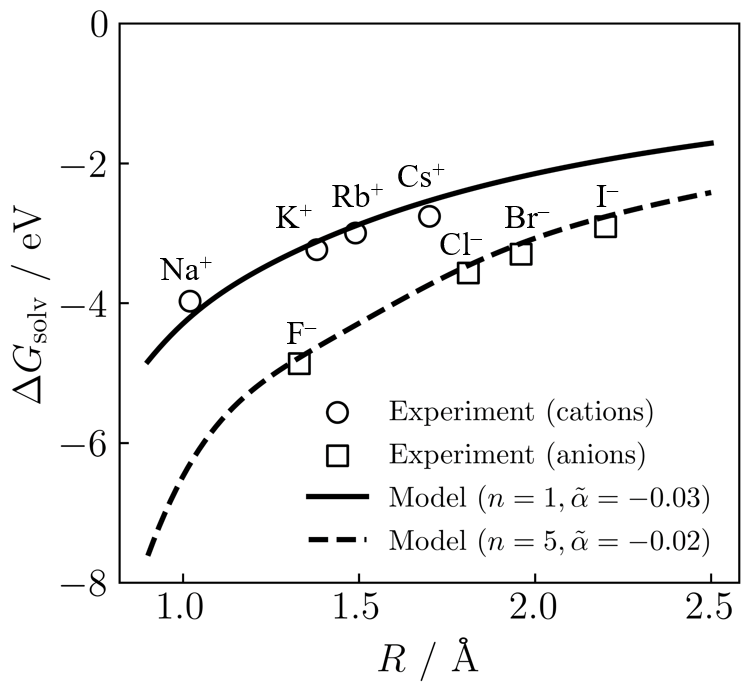}
    \caption{
    Comparisons of experimental hydration energies of alkali metal cations (circles) and halide anions (squares) with those calculated from Eq.~\eqref{equ: solvation energy} using SGn model for the ionic charge distribution. The dimensionless short-range ion--solvent correlation parameter is defined as $\tilde{\alpha}=\alpha\epsilon_0/a_0^2$, with the referenced distance $a_0=1\ \text{\AA}$. Other parameters used are: $K_\alpha = -0.047\ \text{\AA}^2$ (determined from Eq.~\eqref{equ: K_alpha, K_beta}), $K_\beta = 0.0026\ \text{\AA}^4$ (determined from Eq.~\eqref{equ: K_alpha, K_beta}), $\varepsilon_\text{ir}=4.9$ (Ref.~\cite{kornyshev1996shape}), $\varepsilon_s=78.5$ (Ref.~\cite{kornyshev1996shape}). Ionic crystal radii are used for $R$, and their values together with the corresponding hydration energies are taken from Ref.~\cite{marcus2015ions}.  
    }
    \label{fig: comparison of solvation energies}
\end{figure}

Comparisons between the experimental hydration energies of alkali metal cations (circles) and halide anions (squares) and those calculated from Eq.~\eqref{equ: solvation energy} using the SGn model (Eq.~\eqref{equ: SGn model}) for the ionic charge distribution are shown in FIG.~\ref{fig: comparison of solvation energies}. Ionic crystal radii are used for $R$, and their values together with the corresponding hydration energies are taken from Ref.~\cite{marcus2015ions}. Experimental data show that both anions and cations exhibit stronger hydration at smaller ionic radii, a behavior attributed to the enhanced polarization of water by the more intense electric fields of the smaller ions. Moreover, anions are found to be more strongly hydrated than cations at comparable ionic radii, e.g., $\text{K}^+$ compared with $\text{F}^-$, a phenomenon referred to as charge hydration asymmetry\cite{mukhopadhyay2012charge,mukhopadhyay2015accurate}. As seen in FIG.~\ref{fig: comparison of solvation energies}, the monotonic dependence of hydration energies on ionic radius, observed experimentally for alkali metal cations and halide anions, can be quantitatively reproduced by the solvation model in Eq.~\eqref{equ: solvation energy}, provided that different sets of $\alpha$ and $n$ are used for the two ion classes. However, if either the smeared charge effect (e.g., $n=50$) or the short-range ion--solvent correlation effect ($\alpha=0$) is neglected, the charge hydration asymmetry cannot be accounted for, as shown in FIG.~S1 and S2 in Ref.\cite{supplemental}. \rev{As shown in Fig.~S1, the solvation energy exhibits pronounced oscillatory behavior with the ion radius when the charge distribution is localized near the ion surface. In the present model, this oscillation is smoothed by the smeared charge distribution, whereas it can also be smoothed by allowing fluctuations in the ionic (or cavity) radius\cite{basilevsky1998nonlocal}.} 

The contribution of short-range ion--solvent correlations to the solvation energy is given \cite{zhang2025structured} by $\int\text{d}\mathbf{r}\ \alpha \rho(\mathbf{r})\rho_\text{pol}(\mathbf{r})$, where $\rho_\text{pol}$ denotes the solvent polarization charge density. Negative values of $\alpha$ imply that short-range ion--solvent correlations partially cancel the electrostatic interactions between the ionic charge and the solvent polarization charge. This partial cancellation destabilizes the electrostatically ordered solvation structure, thereby weakening solvation. Consequently, a negative $\alpha$ corresponds to a net short-range repulsive interaction between the ion and solvent, whereas a positive $\alpha$ corresponds to a net short-range attractive interaction. A more negative value of $\alpha$, as used for alkali metal cations in FIG.~\ref{fig: comparison of solvation energies}, indicates a stronger net repulsion between the cation and water molecules. In addition, a smaller value of $n$ corresponds to a more broadly distributed ionic charge for cations. Together, these two factors exclude water molecules farther away
from the cationic surface than from the anionic surface. \rev{This can be seen from the ion--water radial distribution functions (RDFs). For example, although $\text{Na}^+$ has a smaller ionic crystal radius than $\text{Cl}^-$, the first peak of the $\text{Na}^+$--$\text{O}$ RDF occurs at a larger distance from the ion center than the first peak of the $\text{Cl}^-$--$\text{H}$ RDF\cite{dang1991ion}. This observation indicates that, relative to its ionic crystal size, water molecules remain farther from the cationic surface.} Therefore, the stronger cation--water repulsion combined with the more smeared charge distribution in our solvation model leads to a larger Born radius for alkali metal cations relative to halide anions, thereby giving rise to charge hydration asymmetry. \rev{The present model provides a phenomenological rationalization of this effect based on a simple solvation formula in Eq.~\eqref{equ: solvation energy}, rather than a detailed microscopic explanation. More specific microscopic origins proposed in the literature include stronger dispersion interactions of the highly polarizable halide anions\cite{duignan2013continuum}, solvent orientation effects\cite{hunenberger2015single}, asymmetry in the charge distribution of water molecules\cite{mobley2008charge}, and differences in the water density within the hydration shells of anions and cations\cite{dinpajooh2015free}.} 

\section{Interfacial solvation}
The solvent structure at electrochemical interfaces is strongly influenced by intricate interactions between the solvent and the electrode surface, encompassing electrostatic forces, van der Waals interactions, and even covalent contributions such as partial charge transfer of water molecules at the Pt(111)--water interface\cite{le2017determining}. Moreover, free electrons spill out from the electrode surface by approximately 1–2 \text{\AA}, which induces a substantial surface potential and markedly affects the electrostatic field sensed by the solvent. Therefore, accurate modeling of the interfacial solvent structure necessitates careful consideration of near-surface physics. The DPPFT approach incorporates these effects within a grand-canonical framework. Specifically, the free electrons of the electrode are treated as an inhomogeneous gas using orbital-free DFT, while the short-range interactions between electrolyte solution particles and the electrode surface are semi-empirically described by Lennard--Jones or Morse potentials\cite{zhang2025structured,huang2023density,huang2021grand}. \rev{The gradient coefficient in the kinetic energy functional and the parameters of the short-range interaction potentials are electrode-specific. They can be calibrated using the recently developed Kohn--Sham--Poisson--Boltzmann model \cite{li2026benchmarking} and DFT-computed binding energies of electrolyte components\cite{tang2023origin}, respectively.} In addition, the Anderson--Newns--Schmickler chemisorption model can be employed to explicitly account for partial charge transfer\cite{huang2024variants}. By solving the closed set of equations derived within DPPFT, one can obtain the spatial distributions of particle densities (including both classical electrolyte solution particles and quantum electrons), the electrostatic potential, and the solvent polarization (see Ref.~\cite{zhang2025structured} or Supporting Materials\cite{supplemental} for details).

\begin{figure}
	\centering
	\includegraphics[width=0.7\linewidth]{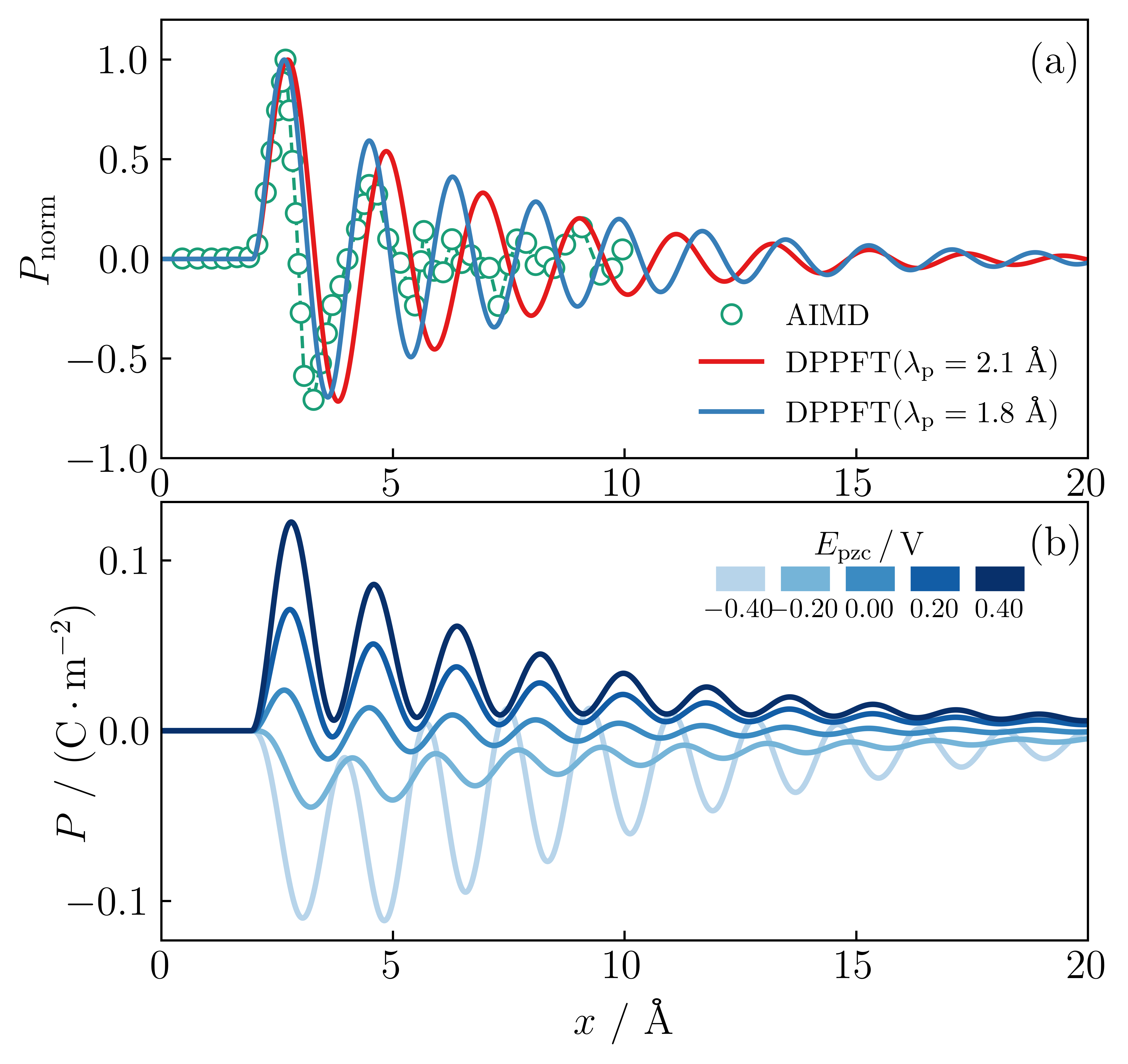}
	\caption{(a) Water polarization at the PZC of the Ag(111)--water interface from AIMD simulations (circles) and from the DPPFT model for the Ag(111)--100 mM NaF aqueous solution interface. In the DPPFT calculations,  $K_\alpha$ and $K_\beta$ are obtained using $\lambda_\text{p}=2.1\ \text{\AA}$ for pure water and an adjusted $\lambda_\text{p}=1.8\ \text{\AA}$ for interfacial water, while the same decay length $\lambda_\text{d}=4.3\ \text{\AA}$ is used in both cases. All other parameters are as listed in Table.~S1 in Ref.\cite{supplemental} (see also Refs.\cite{johnson1972optical,amann2016x} therein). The water polarization is normalized to its maximum value. (b) Water polarization simulated by the DPPFT model with $\lambda_\text{p}=1.8\ \text{\AA}$ at different electrode potentials relative to the PZC, as indicated in the figure. In both panels, the metal surface is located at $x=0$ \text{\AA}.
	}
	\label{fig: interfacial solvation}
\end{figure}

FIG.~\ref{fig: interfacial solvation}a compares the water polarization profiles simulated by ab initio molecular dynamics (AIMD) \cite{le2018structure} and by DPPFT model. AIMD simulations are performed at the Ag(111)--water interface\cite{le2018structure}, while the DPPFT calculations are carried out at the PZC for the Ag(111)--100 mM NaF aqueous solution interface, corresponding to an electrochemical potential of $-3.98$ eV inferred from experimental $C_\text{dl}$ profiles\cite{valette1989double}. For the blue curve in FIG.~\ref{fig: interfacial solvation}a, we use the short-range ion--solvent correlation parameters determined from the ion solvation energies in Section.~\ref{sec: ion solvation}, i.e., $\tilde{\alpha}_c=-0.03$ and $\tilde{\alpha}_a=-0.02$, as well as $K_\alpha$ and $K_\beta$ obtained from the dielectric susceptibility spectrum in Section.~\ref{sec: pure water}, corresponding to a periodic length of 2.1 $\text{\AA}$ and a decay length of 4.3 $\text{\AA}$, as defined in Eqs.~\eqref{equ: periodic length} and \eqref{equ: decay length}. Other parameters used in the DPPFT model are listed in Table.~S1 in Ref.\cite{supplemental}. As shown in FIG.~\ref{fig: interfacial solvation}a, the DPPFT model can reproduce the key features of the water profile observed in AIMD simulations, namely its oscillatory decay toward the bulk solution. The first water layer adopts an oxygen-down configuration \rev{(i.e., an average orientation in which the oxygen atoms of water molecules point toward the electrode surface)}, which is attributed to the positive electric field induced by electron spillover. Subsequent layers exhibit alternating orientations due to hydrogen bonding. The AIMD-predicted water layers are more compact than those obtained from DPPFT using $K_\alpha$ and $K_\beta$ determined from the dielectric response of pure water. This suggests that the periodic length at the Ag(111)--water interface is smaller that that in pure water. From the AIMD results, the interfacial water periodic length can be identified as the separation between the first two polarization peaks, i.e., 1.8 $\text{\AA}$. Assuming the same decay length as in pure water, the short-range correlation parameters $K_\alpha$ and $K_\beta$ for interfacial water can be inversely determined from Eq.~\eqref{equ: periodic length} and Eq.~\eqref{equ: decay length} as follows:
\begin{align}
    K_\alpha = -\frac{(a^2-b^2)\gamma_0}{2(a^2+b^2)^2},\quad K_\beta = \frac{\gamma_0}{4(a^2+b^2)^2},\quad\text{with}\ a=\frac{2\pi}{\lambda_\text{p}},\ b=\frac{1}{\lambda_\text{d}}. \label{equ: alternative determination of Ka and Kb}
\end{align}
With $K_\alpha$ and $K_\beta$ tuned for interfacial water, while all other parameters are kept fixed as listed in Table~S1 in Ref.\cite{supplemental}, the DPPFT model (red curve in FIG.~\ref{fig: interfacial solvation}) achieves improved agreement with the water polarization profile from AIMD simulations. Thus, Eq.~\eqref{equ: alternative determination of Ka and Kb} provides an alternative and more appropriate method for determining $K_\alpha$ and $K_\beta$ at the interface from AIMD simulations.

FIG.~\ref{fig: interfacial solvation}b shows the water polarization profiles near the Ag(111) surface, simulated using the DPPFT model with $\lambda_\text{p} = 1.8~\text{\AA}$ at varying electrode potentials relative to the PZC, denoted as $E_\text{pzc}$. At positive potentials, the first and subsequent water layers adopt an oxygen-down configuration due to the presence of a positive electric field acting on the solvent. As the electrode potential decreases toward negative values, the water layers gradually transition to a hydrogen-down configuration \rev{(i.e., an average orientation in which the hydrogen atoms of water molecules point toward the electrode surface)}. We also observe that the first water layer shifts closer to the metal surface at more negative potentials, consistent with AIMD simulations for the Ag(111)–aqueous solution interface\cite{li2022unraveling}. However, the shift of the water polarization peak toward the metal surface at more positive potentials reported in AIMD studies\cite{li2022unraveling} is not captured by the present DPPFT model. This discrepancy is likely due to the omission of water chemisorption in the current DPPFT model. In our previous work\cite{zhang2025structured}, the orientational transition of the water layer was found to be insensitive to changes in electrode potential; specifically, the first water layer adopted an oxygen-down configuration even at $-0.2~\text{V}$ relative to the PZC. This is attributed to an overestimation of the electron spillover effect in the earlier model, which produced an excessively strong positive electric field acting on solvent molecules near the surface. To mitigate this issue, in the present work we introduce a vacuum region between the metal surface and the first water layer, where the dielectric constant is set to unity. The spatial distribution of the infrared dielectric constant implemented in the DPPFT model is depicted in FIG.~S3. The presence of the vacuum region leads to reduced screening of the surface potential, thereby pushing electrons back into the metal and weakening the electron spillover effect. Naturally, introducing this vacuum region will decrease the magnitude of $C_\text{dl}$ compared with the previous model\cite{zhang2025structured}. To fundamentally avoid the overestimation of electron spillover effect, a more accurate electronic energy functional must be employed. 

\section{Interfacial ionic structure}
\begin{figure}
    \centering
    \includegraphics[width=0.7\linewidth]{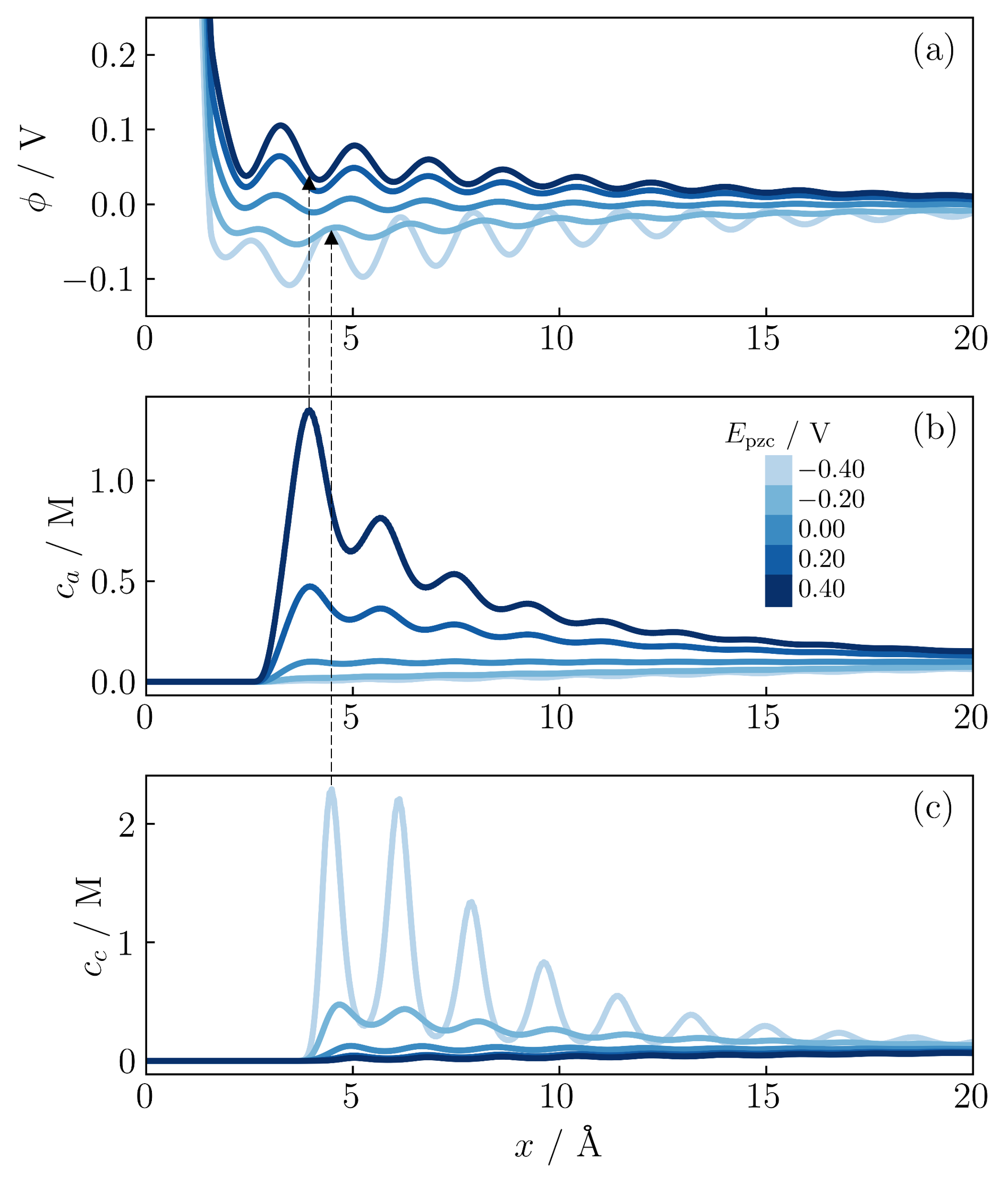}
    \caption{Spatial profiles of (a) water polarization, (b) anionic concentration, and (c) cationic concentration obtained from DPPFT simulations using the parameters in Table~S1 in Ref.\cite{supplemental}, at different electrode potentials relative to the PZC (as indicated in FIG.~\ref{fig: ion distribution}b). The metal surface is located at $x = 0$ \text{\AA} in all panels.}
    \label{fig: ion distribution}
\end{figure}

A characteristic feature of ion distribution near a charged electrode surface is the excess of ions in the diffuse layer, which gradually diminishes toward the bulk solution over a distance characterized by the Debye length. This behavior originates from the electrostatic interactions between the free charges on the electrode surface and ionic charges, as described by Poisson–Boltzmann-type theories. Surface phenomena such as electron spillover and partial charge transfer can, however, alter the surface potential and thereby modulate these electrostatic interactions. FIG.~\ref{fig: ion distribution}b and \ref{fig: ion distribution}c present the concentration profiles of anions and cations obtained from DPPFT simulations. It is evident that counter-ions form layered structures at the interface, with their concentrations exhibiting an oscillatory decay toward the bulk solution. The ionic layering has also been reported in AIMD simulations of the Ag(111)--NaF aqueous solution interface\cite{li2022unraveling}. The emergence of such ionic layering is apparently related to the interactions between ions and interfacial water layers, as shown in FIG.~\ref{fig: interfacial solvation}. On the one hand, oscillatory water polarization gives rise to alternating distributions of polarization charge with opposite signs, i.e., $\rho_\text{pol} = -\nabla \cdot \mathbf{P}$, which in turn generate local minima and maxima in the electrostatic potential on the solution side (FIG.~\ref{fig: ion distribution}a). These potential extrema, corresponding to centers of positive and negative polarization charges, tend to attract anions and cations, respectively, leading to their accumulation near corresponding regions. However, as shown in FIG.~\ref{fig: ion distribution}b and~\ref{fig: ion distribution}c, the peak positions of the anionic and cationic layers do not coincide with the potential minima or maxima. For instance, the first anionic peak at 0.4 V relative to the PZC approximately corresponds to a potential minimum, whereas the first cationic peak at --0.4 V relative to the PZC corresponds to a potential maximum, \rev{as indicated by the dashed arrows in FIG.~\ref{fig: ion distribution}}. This finding indicates that electrostatic interactions between ions and polarized water layers cannot account for the ionic layering observed in the present results.

\begin{figure}
	\centering
	\includegraphics[width=0.75\linewidth]{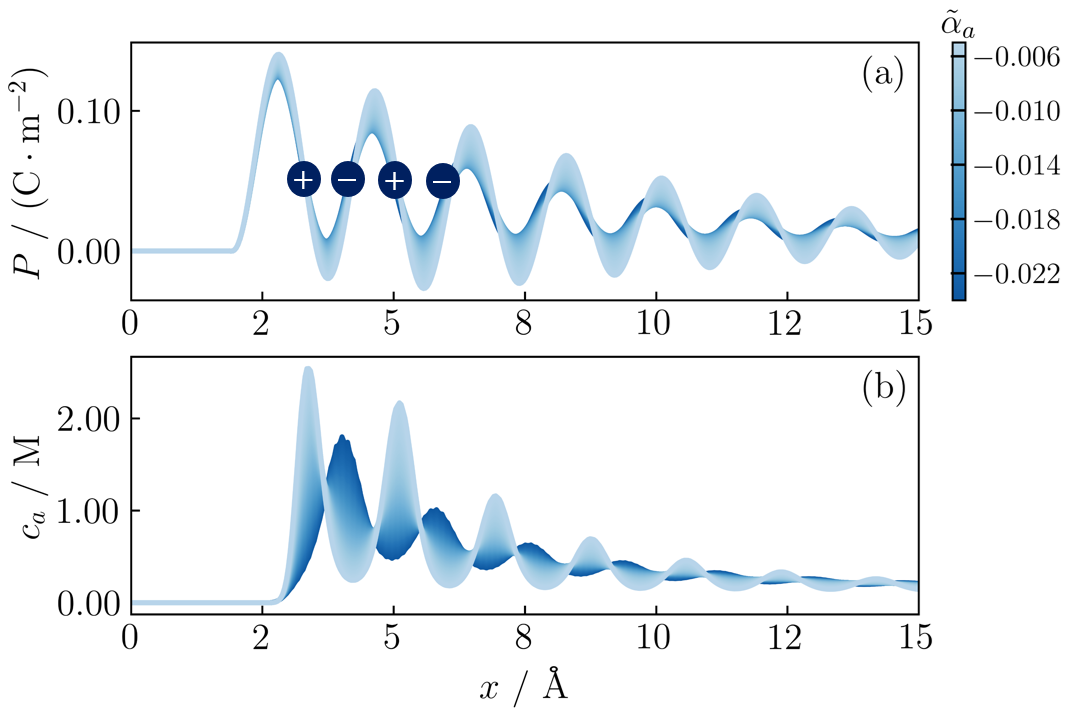}
	\caption{Spatial profiles of (a) water polarization and (b) anionic concentration obtained from DPPFT simulations at an electrode potential of 0.4 V relative to the PZC, for different $\alpha_a$ values as indicated, with all other parameters fixed (Table~S1 in Ref.\cite{supplemental}). The metal surface is located at $x = 0$ \text{\AA} in both panels. The polarization charge centers are indicated in FIG.~\ref{fig: short-range ion-solvent correlation effect}a.}
	\label{fig: short-range ion-solvent correlation effect}
\end{figure}

On the other hand, ions exhibit repulsive short-range correlations with the solvent, characterized by the parameters $\alpha_c$ and $\alpha_a$. As discussed in Section~\ref{sec: ion solvation}, a more negative value of $\alpha_c$ or $\alpha_a$ corresponds to a stronger repulsive effect. FIG.~\ref{fig: short-range ion-solvent correlation effect} shows the profiles of water polarization and anionic concentration at an electrode potential of 0.4 V relative to the PZC, under different $\alpha_a$ values while keeping all other parameters fixed, as listed in Table~S1 in Ref.\cite{supplemental}. When $\alpha_a \approx 0$, i.e., in the absence of short-range ion–solvent correlations, anions are attracted to regions near the centers of positive polarization charges, as shown in FIG.~\ref{fig: short-range ion-solvent correlation effect}b. As $\alpha_a$ becomes more negative, the electric susceptibility of water decreases (Eq.~\eqref{equ: chi_alpha}), reflecting a reduced polarization response of each water layer (FIG.~\ref{fig: short-range ion-solvent correlation effect}a). Consequently, the magnitude of the positive polarization charge decreases, weakening its attraction to anions. In addition, a more negative $\alpha_a$ implies a stronger repulsive interaction between the anionic charge and the positive polarization charges, i.e., the hydrogen ends of water molecules. This repulsion pushes the anions into the regions between adjacent layers of positive polarization charges, corresponding to the centers of negative polarization charges (FIG.~\ref{fig: short-range ion-solvent correlation effect}b). This behavior suggests that anions tend to reside within water layers in order to preserve a solvation configuration similar to that in bulk solution, namely with the hydrogen ends (positive polarization charges) of water molecules oriented toward the anion. Moreover, a more negative $\alpha$ for cations than for anions indicates that cations experience a stronger repulsive effect from water molecules, thereby forming more distinct interfacial layering. Such behavior is also observed in AIMD simulations \cite{li2022unraveling}.

\rev{Ionic layering has also often been observed at highly concentrated electrolyte interfaces\cite{philpott1995screening,spohr1999molecular,de2020interfacial,mezger2008molecular}, whose limiting case corresponds to ionic liquid interfaces. However, the physical origin of ionic layering at highly concentrated electrolyte interfaces is fundamentally different from that at the dilute electrolyte interfaces considered in the present work. At dilute electrolyte interfaces, the solvent is the dominant component and forms a layered structure due to short-range solvent--solvent correlations, as discussed above. Ionic layering then arises from the interactions of ions with these solvent layers. In particular, short-range ion--solvent correlations influence how the ions fit within the solvent layers, as shown in FIG.~\ref{fig: short-range ion-solvent correlation effect}. In contrast, at highly concentrated electrolyte interfaces, ions become the dominant component, and the ionic layering originates from short-range ion--ion correlations. At moderately concentrated electrolyte interfaces, the interfacial structure is expected to transition from being governed by short-range solvent--solvent correlations to short-range ion--ion correlations as the concentration increases. Consequently, at moderately concentrated electrolyte interfaces, short-range solvent--solvent, ion--solvent, and ion--ion correlations all play important roles in determining the interfacial structure.} 

\rev{The short-range ion–ion correlations are not accounted for in the present DPPFT model. Therefore, the applicability of the present DPPFT model is limited to electrolyte concentrations below a certain threshold. Recent molecular dynamics simulations by Chen et al.\cite{chen2025nonlocal} show that the peak positions in the oscillatory part of the ion–ion correlation functions remain largely independent of concentration up to 1.9 M, indicating that the electrolyte structure is fundamentally governed by water. This observation suggests that the present DPPFT model remains applicable for electrolyte concentrations up to approximately 1.9 M. A natural direction for future development of the DPPFT model is to overcome this limitation and extend its applicability across the full range of electrolyte concentrations. This can be achieved by incorporating short-range ion--ion correlations into the field theory of electrolyte solutions, with appropriate choices of nonlocal correlation kernels. Some theoretical models\cite{bazant2011double,blossey2017structural,gavish2016theory,limmer2015interfacial} have been developed in this regard that reproduce interfacial layering at ionic liquid interfaces. Furthermore, the correlation functions or structure factors can be obtained within the random phase (one-loop) approximation from the field-theoretic formulation\cite{dinpajooh2024beyond,becker2025dielectric,weyman2022direct}. By comparing them with those computed from molecular dynamics simulations, the short-range correlation parameters can be self-consistently determined.} 

\rev{It would also be of considerable interest to combine the DPPFT model with electron transfer theory to describe reactive electrochemical interfaces. One of the authors of the present work has previously developed a unified theoretical framework that integrates continuum EDL models with the Anderson--Newns--Schmickler theory of chemisorption\cite{huang2020mixed}. This framework enables the calculation of key electron-transfer parameters, such as the solvent reorganization energy, reaction free energy, and activation free energy, while also providing access to local reaction conditions, including the local electrostatic potential, reactant concentration, and dielectric response. A promising direction for future work is to refine this framework by employing the DPPFT model to describe the EDL. An interesting question is how short-range correlations and interfacial layering influence key electron-transfer parameters and local reaction conditions, and thereby affect electron transfer kinetics.}              

\section{Conclusions}
In this work, we first discussed a scheme to determine the short-range correlation parameters in recently developed field-theoretic approach\cite{blossey2022field,blossey2023comprehensive} for structured electrolyte solutions. Within this framework, solvent–solvent correlations are described by a fourth-order Landau–Ginzburg (LG) functional, which reproduces the primary peak in the wave-number-dependent dielectric susceptibility spectrum of pure water. The relevant parameters can therefore be determined from the peak position. The dielectric susceptibility associated with the LG functional takes a Yukawa-type form with a decay length, multiplied by an oscillatory term characterized by a periodic length. On this basis, the ion solvation energy in the linear response regime is derived from nonlocal electrostatics. The short-range ion–solvent correlation parameters for alkali metal cations and halide anions are then obtained by fitting the experimental ionic-radius-dependent hydration energies. The more negative value found for cation–solvent interactions indicates stronger short-range repulsion, thereby rationalizing the observed charge hydration asymmetry.

With the short-range correlation parameters determined from pure water and ion solvation, the interfacial structure of the Ag(111)–NaF aqueous electrolyte system was investigated within the density–potential–polarization functional theory (DPPFT). Water polarization profiles from AIMD simulations are more compact than those from the initial DPPFT calculations, suggesting a shorter periodic length for interfacial water. When solvent–solvent correlation parameters corresponding to this reduced periodic length are adopted, the DPPFT model shows improved agreement with AIMD results. Furthermore, the interactions between ions and interfacial water layers results in the formation of oscillatory ionic layering. As the strength of short-range ion–solvent repulsion increases, the peaks of anionic/cationic layers shift from regions near the centers of positive/negative polarization charges toward those of negative/positive polarization charges, in order to preserve solvation configurations similar to those in bulk solution. In conclusion, we have established a practical methodology for determining short-range correlation parameters within the DPPFT framework. This semi-classical and computationally efficient approach enables a quantitative description of short-range correlation effects and provides a refined understanding of the interfacial structure of macroscopic electrochemical systems.

\begin{acknowledgments}
	M.Z. and J.H. are supported by European Research Council (ERC) Starting Grant (MESO-CAT, Grant agreement No. 101163405).
\end{acknowledgments}

\bibliographystyle{unsrt}
\bibliography{references}
\end{document}

%% file: preamble.tex
\usepackage{amsmath,amssymb,bm}
\usepackage{graphicx}
\usepackage{mathrsfs}  
\usepackage{makecell}
\usepackage{enumitem}
\usepackage{mdframed}
\usepackage{xcolor}     
\newcommand{\abs}[1]{\left\lvert #1 \right\rvert}
\usepackage{empheq}
\usepackage{setspace}
\usepackage[colorlinks=true,linkcolor=blue,citecolor=blue,urlcolor=blue]{hyperref}
 

%% file: references.bib
@article{basilevsky1998nonlocal,
	title={Nonlocal continuum solvation model with oscillating susceptibility kernels: {A} nonrigid cavity model},
	author={Basilevsky, Mikhail V and Parsons, Drew F},
	journal={The Journal of chemical physics},
	volume={108},
	number={21},
	pages={9114--9123},
	year={1998},
	publisher={American Institute of Physics}
}

@article{dinpajooh2015free,
	title={Free energy of ion hydration: {I}nterface susceptibility and scaling with the ion size},
	author={Dinpajooh, Mohammadhasan and Matyushov, Dmitry V},
	journal={The Journal of Chemical Physics},
	volume={143},
	number={4},
	year={2015},
	publisher={AIP Publishing}
}

@misc{supplemental,
	note = {See Supplemental Material at [URL will be inserted by publisher] for derivations of the solvent dielectric susceptibility and ion solvation energy, figures illustrating the effects of charge smearing and short-range ion--solvent correlations on ion solvation energy, and details of the implementation and model parameters of the DPPFT model.}
}

@article{huang2020mixed,
	title={Mixed quantum-classical treatment of electron transfer at electrocatalytic interfaces: {T}heoretical framework and conceptual analysis},
	author={Huang, Jun},
	journal={The Journal of Chemical Physics},
	volume={153},
	number={16},
	year={2020},
	publisher={AIP Publishing}
}

@article{weyman2022direct,
	title={Direct calculation of the functional inverse of realistic interatomic potentials in field-theoretic simulations},
	author={Weyman, Alexander and Mavrantzas, Vlasis G and {\"O}ttinger, Hans Christian},
	journal={The Journal of Chemical Physics},
	volume={156},
	number={22},
	year={2022},
	publisher={AIP Publishing}
}

@article{dinpajooh2024beyond,
	title={Beyond the Debye--H{\"u}ckel limit: Toward a general theory for concentrated electrolytes},
	author={Dinpajooh, Mohammadhasan and Intan, Nadia N and Duignan, Timothy T and Biasin, Elisa and Fulton, John L and Kathmann, Shawn M and Schenter, Gregory K and Mundy, Christopher J},
	journal={The Journal of Chemical Physics},
	volume={161},
	number={23},
	year={2024},
	publisher={AIP Publishing}
}

@article{limmer2015interfacial,
	title={Interfacial ordering and accompanying divergent capacitance at ionic liquid-metal interfaces},
	author={Limmer, David T},
	journal={Physical review letters},
	volume={115},
	number={25},
	pages={256102},
	year={2015},
	publisher={APS}
}

@article{gavish2016theory,
	title={Theory of phase separation and polarization for pure ionic liquids},
	author={Gavish, Nir and Yochelis, Arik},
	journal={The journal of physical chemistry letters},
	volume={7},
	number={7},
	pages={1121--1126},
	year={2016},
	publisher={ACS Publications}
}

@article{blossey2017structural,
	title={Structural interactions in ionic liquids linked to higher-order Poisson-Boltzmann equations},
	author={Blossey, Ralf and Maggs, AC and Podgornik, R},
	journal={Physical Review E},
	volume={95},
	number={6},
	pages={060602},
	year={2017},
	publisher={APS}
}

@article{bazant2011double,
	title={Double layer in ionic liquids: {O}verscreening versus crowding},
	author={Bazant, Martin Z and Storey, Brian D and Kornyshev, Alexei A},
	journal={Physical review letters},
	volume={106},
	number={4},
	pages={046102},
	year={2011},
	publisher={APS}
}

@article{chen2025nonlocal,
	title={Nonlocal dielectric response of aqueous electrolytes and decay behavior of ionic correlations},
	author={Chen, Ming and Feng, Guang and Kjellander, Roland and Kornyshev, Alexei A},
	journal={Physical Review E},
	volume={112},
	number={6},
	pages={065407},
	year={2025},
	publisher={APS}
}

@article{mezger2008molecular,
	title={Molecular layering of fluorinated ionic liquids at a charged sapphire (0001) surface},
	author={Mezger, Markus and Schroder, Heiko and Reichert, Harald and Schramm, Sebastian and Okasinski, John S and Schoder, Sebastian and Honkimaki, Veijo and Deutsch, Moshe and Ocko, Benjamin M and Ralston, John and others},
	journal={Science},
	volume={322},
	number={5900},
	pages={424--428},
	year={2008},
	publisher={American Association for the Advancement of Science}
}

@article{de2020interfacial,
	title={Interfacial layering in the electric double layer of ionic liquids},
	author={de Souza, J Pedro and Goodwin, Zachary AH and McEldrew, Michael and Kornyshev, Alexei A and Bazant, Martin Z},
	journal={Physical Review Letters},
	volume={125},
	number={11},
	pages={116001},
	year={2020},
	publisher={APS}
}

@article{tang2023origin,
	title={Origin of solvent dependency of the potential of zero charge},
	author={Tang, Weiqiang and Zhao, Shuangliang and Huang, Jun},
	journal={JACS Au},
	volume={3},
	number={12},
	pages={3381--3390},
	year={2023},
	publisher={ACS Publications}
}

@article{li2026benchmarking,
	title={Benchmarking orbital-free density-potential functional theory of electrified metal-solution interfaces},
	author={Li, Chenkun and Wang, Xiwei and Eikerling, Michael and Huang, Jun},
	journal={Journal of Chemical Theory and Computation},
	volume={22},
	number={5},
	pages={2607--2621},
	year={2026},
	publisher={ACS Publications}
}

@article{mobley2008charge,
	title={Charge asymmetries in hydration of polar solutes},
	author={Mobley, David L and Barber, Alan E and Fennell, Christopher J and Dill, Ken A},
	journal={The journal of physical chemistry B},
	volume={112},
	number={8},
	pages={2405--2414},
	year={2008},
	publisher={ACS Publications}
}

@book{hunenberger2015single,
	title={Single-ion solvation: experimental and theoretical approaches to elusive thermodynamic quantities},
	author={Hunenberger, Philippe and Reif, Maria},
	year={2015},
	publisher={Royal Society of Chemistry}
}

@article{duignan2013continuum,
	title={A continuum model of solvation energies including electrostatic, dispersion, and cavity contributions},
	author={Duignan, Timothy T and Parsons, Drew F and Ninham, Barry W},
	journal={The Journal of Physical Chemistry B},
	volume={117},
	number={32},
	pages={9421--9429},
	year={2013},
	publisher={ACS Publications}
}

@article{dang1991ion,
	title={Ion solvation in polarizable water: {M}olecular dynamics simulations},
	author={Dang, Liem X and Rice, Julia E and Caldwell, James and Kollman, Peter A},
	journal={Journal of the American Chemical Society},
	volume={113},
	number={7},
	pages={2481--2486},
	year={1991},
	publisher={ACS Publications}
}

@article{zhang2025electrochemical,
	title={Electrochemical electron transfer: {K}ey concepts, theories, and parameterization via atomistic simulations},
	author={Zhang, Mengke and Chen, Yanxia and Melander, Marko M and Huang, Jun},
	journal={arXiv preprint arXiv:2510.24635},
	year={2025}
}

@article{ballenegger2005dielectric,
	title={Dielectric permittivity profiles of confined polar fluids},
	author={Ballenegger, V and Hansen, J-P},
	journal={The Journal of chemical physics},
	volume={122},
	number={11},
	year={2005},
	publisher={AIP Publishing}
}

@article{mondal2021anomalous,
	title={Anomalous dielectric response of nanoconfined water},
	author={Mondal, Sayantan and Bagchi, Biman},
	journal={The Journal of Chemical Physics},
	volume={154},
	number={4},
	year={2021},
	publisher={AIP Publishing}
}

@article{dinpajooh2016dielectric,
	title={Dielectric constant of water in the interface},
	author={Dinpajooh, Mohammadhasan and Matyushov, Dmitry V},
	journal={The Journal of Chemical Physics},
	volume={145},
	number={1},
	year={2016},
	publisher={AIP Publishing}
}

@article{schlaich2016water,
	title={Water dielectric effects in planar confinement},
	author={Schlaich, Alexander and Knapp, Ernst W and Netz, Roland R},
	journal={Physical review letters},
	volume={117},
	number={4},
	pages={048001},
	year={2016},
	publisher={APS}
}

@article{fumagalli2018anomalously,
	title={Anomalously low dielectric constant of confined water},
	author={Fumagalli, Laura and Esfandiar, Ali and Fabregas, Rene and Hu, Sheng and Ares, Pablo and Janardanan, Amritha and Yang, Qian and Radha, Boya and Taniguchi, Takashi and Watanabe, Kenji and others},
	journal={Science},
	volume={360},
	number={6395},
	pages={1339--1342},
	year={2018},
	publisher={American Association for the Advancement of Science}
}

@article{mukhopadhyay2012charge,
	title={Charge hydration asymmetry: {The} basic principle and how to use it to test and improve water models},
	author={Mukhopadhyay, Abhishek and Fenley, Andrew T and Tolokh, Igor S and Onufriev, Alexey V},
	journal={The journal of physical chemistry B},
	volume={116},
	number={32},
	pages={9776--9783},
	year={2012},
	publisher={ACS Publications}
}

@article{clementi1974roothaan,
	title={Roothaan-Hartree-Fock atomic wavefunctions: Basis functions and their coefficients for ground and certain excited states of neutral and ionized atoms, Z$\le$54},
	author={Clementi, Enrico and Roetti, Carla},
	journal={Atomic data and nuclear data tables},
	volume={14},
	number={3-4},
	pages={177--478},
	year={1974},
	publisher={Elsevier}
}

@article{gongadze2012decrease,
	title={Decrease of permittivity of an electrolyte solution near a charged surface due to saturation and excluded volume effects},
	author={Gongadze, Ekaterina and Igli{\v{c}}, Ale{\v{s}}},
	journal={Bioelectrochemistry},
	volume={87},
	pages={199--203},
	year={2012},
	publisher={Elsevier}
}

@article{amann2016x,
  title={X-ray and neutron scattering of water},
  author={Amann-Winkel, Katrin and Bellissent-Funel, Marie-Claire and Bove, Livia E and Loerting, Thomas and Nilsson, Anders and Paciaroni, Alessandro and Schlesinger, Daniel and Skinner, Lawrie},
  journal={Chemical reviews},
  volume={116},
  number={13},
  pages={7570--7589},
  year={2016},
  publisher={ACS Publications}
}

@article{johnson1972optical,
  title={Optical constants of the noble metals},
  author={Johnson, Peter B and Christy, R W},
  journal={Physical review B},
  volume={6},
  number={12},
  pages={4370},
  year={1972},
  publisher={APS}
}

@article{huang2021grand,
  title={Grand-canonical model of electrochemical double layers from a hybrid density--potential functional},
  author={Huang, Jun and Chen, Shengli and Eikerling, Michael},
  journal={Journal of chemical theory and computation},
  volume={17},
  number={4},
  pages={2417--2430},
  year={2021},
  publisher={ACS Publications}
}

@article{huang2024variants,
  title={Variants of surface charges and capacitances in electrocatalysis: Insights from density-potential functional theory embedded with an implicit chemisorption model},
  author={Huang, Jun and Dom{\'\i}nguez-Flores, Fabiola and Melander, Marko},
  journal={PRX Energy},
  volume={3},
  number={4},
  pages={043008},
  year={2024},
  publisher={APS}
}

@article{le2017determining,
  title={Determining potentials of zero charge of metal electrodes versus the standard hydrogen electrode from density-functional-theory-based molecular dynamics},
  author={Le, Jiabo and Iannuzzi, Marcella and Cuesta, Angel and Cheng, Jun},
  journal={Physical review letters},
  volume={119},
  number={1},
  pages={016801},
  year={2017},
  publisher={APS}
}

@article{li2022unraveling,
  title={Unraveling molecular structures and ion effects of electric double layers at metal water interfaces},
  author={Li, Lang and Liu, Yun-Pei and Le, Jia-Bo and Cheng, Jun},
  journal={Cell Reports Physical Science},
  volume={3},
  number={2},
  year={2022},
  publisher={Elsevier}
}

@article{le2018structure,
  title={The structure of metal-water interface at the potential of zero charge from density functional theory-based molecular dynamics},
  author={Le, Jiabo and Cuesta, Angel and Cheng, Jun},
  journal={Journal of Electroanalytical Chemistry},
  volume={819},
  pages={87--94},
  year={2018},
  publisher={Elsevier}
}

@article{valette1989double,
  title={Double layer on silver single crystal electrodes in contact with electrolytes having anions which are slightly specifically adsorbed: Part III. The (111) face},
  author={Valette, Georges},
  journal={Journal of electroanalytical chemistry and interfacial electrochemistry},
  volume={269},
  number={1},
  pages={191--203},
  year={1989},
  publisher={Elsevier}
}

@article{mukhopadhyay2015accurate,
  title={Accurate evaluation of charge asymmetry in aqueous solvation},
  author={Mukhopadhyay, Abhishek and Tolokh, Igor S and Onufriev, Alexey V},
  journal={The Journal of Physical Chemistry B},
  volume={119},
  number={20},
  pages={6092--6100},
  year={2015},
  publisher={ACS Publications}
}

@book{marcus2015ions,
  title={Ions in Solution and their Solvation},
  author={Marcus, Yizhak},
  year={2015},
  publisher={John Wiley \& Sons}
}

@article{kornyshev1997nonlocal,
  title={Nonlocal dielectric response of water and reorganization energies for outer sphere electron transfer reactions},
  author={Kornyshev, Alexei A and Sutmann, Godehard},
  journal={Electrochimica acta},
  volume={42},
  number={18},
  pages={2801--2808},
  year={1997},
  publisher={Elsevier}
}

@article{blossey2022continuum,
  title={Continuum theories of structured dielectrics},
  author={Blossey, Ralf and Podgornik, Rudolf},
  journal={Europhysics Letters},
  volume={139},
  number={2},
  pages={27002},
  year={2022},
  publisher={IOP Publishing}
}

@article{huang2023density,
  title={Density-potential functional theory of electrochemical double layers: {Calibration} on the {Ag}(111)-{KPF}$_6$ system and parametric analysis},
  author={Huang, Jun},
  journal={Journal of chemical theory and computation},
  volume={19},
  number={3},
  pages={1003--1013},
  year={2023},
  publisher={ACS Publications}
}

@article{berthoumieux2019dielectric,
  title={Dielectric response in the vicinity of an ion: A nonlocal and nonlinear model of the dielectric properties of water},
  author={Berthoumieux, Helene and Paillusson, Fabien},
  journal={The Journal of Chemical Physics},
  volume={150},
  number={9},
  year={2019},
  publisher={AIP Publishing}
}

@article{berthoumieux2018gaussian,
  title={Gaussian field model for polar fluids as a function of density and polarization: Toward a model for water},
  author={Berthoumieux, H{\'e}l{\`e}ne},
  journal={The Journal of Chemical Physics},
  volume={148},
  number={10},
  year={2018},
  publisher={AIP Publishing}
}

@article{spohr1999molecular,
  title={Molecular simulation of the electrochemical double layer},
  author={Spohr, Eckhard},
  journal={Electrochimica Acta},
  volume={44},
  number={11},
  pages={1697--1705},
  year={1999},
  publisher={Elsevier}
}

@article{philpott1995screening,
  title={Screening of charged electrodes in aqueous electrolytes},
  author={Philpott, Michael R and Glosli, James N},
  journal={Journal of the Electrochemical Society},
  volume={142},
  number={2},
  pages={L25},
  year={1995},
  publisher={IOP Publishing}
}

@article{di2023constant,
  title={Constant chemical potential--quantum mechanical--molecular dynamics simulations of the graphene--electrolyte double layer},
  author={Di Pasquale, Nicodemo and Finney, Aaron R and Elliott, Joshua D and Carbone, Paola and Salvalaglio, Matteo},
  journal={The Journal of chemical physics},
  volume={158},
  number={13},
  year={2023},
  publisher={AIP Publishing}
}

@article{martin2016atomically,
  title={Atomically resolved three-dimensional structures of electrolyte aqueous solutions near a solid surface},
  author={Martin-Jimenez, Daniel and Chacon, Enrique and Tarazona, Pedro and Garcia, Ricardo},
  journal={Nature communications},
  volume={7},
  number={1},
  pages={12164},
  year={2016},
  publisher={Nature Publishing Group UK London}
}

@article{becker2025dielectric,
  title={Dielectric properties of aqueous electrolytes at the nanoscale},
  author={Becker, Maximilian R and Netz, Roland R and Loche, Philip and Bonthuis, Douwe Jan and Mouhanna, Dominique and Berthoumieux, H{\'e}l{\`e}ne},
  journal={Physical Review Letters},
  volume={134},
  number={15},
  pages={158001},
  year={2025},
  publisher={APS}
}

@article{labavic2025nonlocal,
  title={Nonlocal dielectric properties of water: the role of electronic delocalisation},
  author={Labavi{\'c}, Darka and Br{\"u}nig, Florian N and Netz, Roland R and Bocquet, Marie-Laure and Berthoumieux, H{\'e}l{\`e}ne},
  journal={arXiv preprint arXiv:2505.11101},
  year={2025}
}

@article{marvcelja1976repulsion,
  title={Repulsion of interfaces due to boundary water},
  author={Mar{\v{c}}elja, S and Radi{\'c}, N},
  journal={Chemical Physics Letters},
  volume={42},
  number={1},
  pages={129--130},
  year={1976},
  publisher={North-Holland}
}

@article{hildebrandt2004novel,
  title={Novel formulation of nonlocal electrostatics},
  author={Hildebrandt, Andreas and Blossey, Ralf and Rjasanow, Sergej and Kohlbacher, Oliver and Lenhof, H-P},
  journal={Physical review letters},
  volume={93},
  number={10},
  pages={108104},
  year={2004},
  publisher={APS}
}

@article{kornyshev1983non,
  title={Non-local dielectric response of a polar solvent and {Debye} screening in ionic solution},
  author={Kornyshev, Alexei A},
  journal={Journal of the Chemical Society, Faraday Transactions 2: Molecular and Chemical Physics},
  volume={79},
  number={5},
  pages={651--661},
  year={1983},
  publisher={Royal Society of Chemistry}
}

@article{kornyshev1982nonlocal,
  title={Nonlocal electrostatic approach to the problem of a double layer at a metal-electrolyte interface},
  author={Kornyshev, AA and Schmickler, W and Vorotyntsev, MA},
  journal={Physical Review B},
  volume={25},
  number={8},
  pages={5244},
  year={1982},
  publisher={APS}
}

@article{kornyshev1981nonlocaldoublelayer,
  title={Nonlocal dielectric response of the electrode/solvent interface in the double layer problem},
  author={Kornyshev, AA and Vorotyntsev, MA},
  journal={Canadian Journal of Chemistry},
  volume={59},
  number={13},
  pages={2031--2042},
  year={1981},
  publisher={NRC Research Press Ottawa, Canada}
}

@incollection{ulstrup1979multiphonon,
  title={Multiphonon Representation of Continuous Media},
  author={Ulstrup, Jens},
  booktitle={Charge Transfer Processes in Condensed Media},
  pages={40--70},
  year={1979},
  publisher={Springer}
}

@article{gongadze2015asymmetric,
  title={Asymmetric size of ions and orientational ordering of water dipoles in electric double layer model-an analytical mean-field approach},
  author={Gongadze, Ekaterina and Igli{\v{c}}, Ale{\v{s}}},
  journal={Electrochimica Acta},
  volume={178},
  pages={541--545},
  year={2015},
  publisher={Elsevier}
}

@article{gongadze2014ions,
  title={Ions and water molecules in an electrolyte solution in contact with charged and dipolar surfaces},
  author={Gongadze, Ekaterina and Velikonja, Alja{\v{z}} and Perutkova, {\v{S}}arka and Kramar, Peter and Ma{\v{c}}ek-Lebar, Alenka and Kralj-Igli{\v{c}}, Veronika and Igli{\v{c}}, Ale{\v{s}}},
  journal={Electrochimica Acta},
  volume={126},
  pages={42--60},
  year={2014},
  publisher={Elsevier}
}

@article{kornyshev1996shape,
  title={The shape of the nonlocal dielectric function of polar liquids and the implications for thermodynamic properties of electrolytes: A comparative study},
  author={Kornyshev, Alexei A and Sutmann, Godehard},
  journal={The Journal of chemical physics},
  volume={104},
  number={4},
  pages={1524--1544},
  year={1996},
  publisher={American Institute of Physics}
}

@article{dogonadze1974polar,
  title={Polar solvent structure in the theory of ionic solvation},
  author={Dogonadze, RR and Kornyshev, AA},
  journal={Journal of the Chemical Society, Faraday Transactions 2: Molecular and Chemical Physics},
  volume={70},
  pages={1121--1132},
  year={1974},
  publisher={Royal Society of Chemistry}
}

@article{bopp1996static,
  title={Static nonlocal dielectric function of liquid water},
  author={Bopp, Philippe A and Kornyshev, Alexei A and Sutmann, Godehard},
  journal={Physical review letters},
  volume={76},
  number={8},
  pages={1280},
  year={1996},
  publisher={APS}
}

@article{kornyshev1981nonlocal,
  title={Nonlocal screening of ions in a structurized polar liquid—new aspects of solvent description in electrolyte theory},
  author={Kornyshev, AA},
  journal={Electrochimica Acta},
  volume={26},
  number={1},
  pages={1--20},
  year={1981},
  publisher={Elsevier}
}

@book{kuznetsov1999electron,
  title={Electron transfer in chemistry and biology: an introduction to the theory},
  author={Kuznetsov, Alexander M and Ulstrup, Jens},
  year={1999},
  publisher={John Wiley \& Sons Ltd}
}

@article{bopp1998frequency,
  title={Frequency and wave-vector dependent dielectric function of water: Collective modes and relaxation spectra},
  author={Bopp, Philippe A and Kornyshev, Alexei A and Sutmann, Godehard},
  journal={The Journal of chemical physics},
  volume={109},
  number={5},
  pages={1939--1958},
  year={1998},
  publisher={American Institute of Physics}
}

@article{zhang2025structured,
  title={Structured solvent on a split electron tail: A semiclassical theory of electrified metal-solution interfaces},
  author={Zhang, Mengke and Chen, Yanxia and Eikerling, Michael and Huang, Jun},
  journal={Physical Review Applied},
  volume={23},
  number={2},
  pages={024009},
  year={2025},
  publisher={APS}
}

@article{monet2021nonlocal,
  title={Nonlocal dielectric response of water in nanoconfinement},
  author={Monet, Geoffrey and Bresme, Fernando and Kornyshev, Alexei and Berthoumieux, H{\'e}l{\`e}ne},
  journal={Physical Review Letters},
  volume={126},
  number={21},
  pages={216001},
  year={2021},
  publisher={APS}
}

@article{hedley2023dramatic,
  title={The dramatic effect of water structure on hydration forces and the electrical double layer},
  author={Hedley, Jonathan G and Berthoumieux, H{\'e}l{\`e}ne and Kornyshev, Alexei A},
  journal={The Journal of Physical Chemistry C},
  volume={127},
  number={18},
  pages={8429--8447},
  year={2023},
  publisher={ACS Publications}
}

@article{blossey2023comprehensive,
  title={A comprehensive continuum theory of structured liquids},
  author={Blossey, Ralf and Podgornik, Rudolf},
  journal={Journal of Physics A: Mathematical and Theoretical},
  volume={56},
  number={2},
  pages={025002},
  year={2023},
  publisher={IOP Publishing}
}

@article{blossey2022field,
  title={Field theory of structured liquid dielectrics},
  author={Blossey, Ralf and Podgornik, Rudolf},
  journal={Physical Review Research},
  volume={4},
  number={2},
  pages={023033},
  year={2022},
  publisher={APS}
}
